\documentclass[sigconf,natbib=true,nonacm]{acmart}
\usepackage{balance}
\usepackage{todonotes}
\usepackage{multicol}
\usepackage{multirow}
\usepackage[htt]{hyphenat}
\usepackage{pifont}% http://ctan.org/pkg/pifont
\newcommand{\cmark}{\ding{51}}%
\newcommand{\xmark}{\ding{55}}%
\newcolumntype{P}[1]{>{\raggedright\arraybackslash}p{#1}}
\begin{document}

% \title{Impact of Retweets and Other Social Engagement\\on Online Toxicity}
% \title{Social-LLM: Scalable Social Network User Modeling Using Language Features and Network Homophily}
% \title{Social-LLM: Scalable User Social and Behavioral Modeling Using Language and Network Information}
% \title{Social-LLM: Scalable Modeling of User Social Behavior with Language Models and Network Data}
% \title{Social-LLM: Modeling  User Social Behavior at Scale \\using Language Models and Network Data}
\title[Social-LLM: Modeling  User Behavior at Scale using Language Models and Social Network Data]{Social-LLM: Modeling  User Behavior at Scale \\using Language Models and Social Network Data}

\author{Julie Jiang}
\affiliation{%
 \institution{University of Southern California}
 \country{Los Angeles, CA, USA}}
\email{juliej@isi.edu}
\author{Emilio Ferrara}
\affiliation{%
 \institution{University of Southern California}
 \country{Los Angeles, CA, USA}}
\email{emiliofe@usc.edu}

%    <concept>
%        <concept_id>10010405.10010455.10010461</concept_id>
%        <concept_desc>Applied computing~Sociology</concept_desc>
%        <concept_significance>500</concept_significance>
%        </concept>
%    <concept>
%        <concept_id>10003120.10003130.10011762</concept_id>
%        <concept_desc>Human-centered computing~Empirical studies in collaborative and social computing</concept_desc>
%        <concept_significance>500</concept_significance>
%        </concept>
%    <concept>
%        <concept_id>10010147.10010257</concept_id>
%        <concept_desc>Computing methodologies~Machine learning</concept_desc>
%        <concept_significance>300</concept_significance>
%        </concept>
%  </ccs2012>

% \ccsdesc[500]{Applied computing~Sociology}
% \ccsdesc[500]{Human-centered computing~Empirical studies in collaborative and social computing}
% \ccsdesc[300]{Computing methodologies~Machine learning}
\keywords{Social Network, User Detection, User Behavior, Network Homophily}

\begin{abstract}
The proliferation of social network data has unlocked unprecedented opportunities for extensive, data-driven exploration of human behavior. The structural intricacies of social networks offer insights into various computational social science issues, particularly concerning social influence and information diffusion. However, modeling large-scale social network data comes with computational challenges. Though large language models make it easier than ever to model textual content, any advanced network representation methods struggle with scalability and efficient deployment to out-of-sample users. In response, we introduce a novel approach tailored for modeling social network data in user detection tasks. This innovative method integrates localized social network interactions with the capabilities of large language models. Operating under the premise of social network homophily, which posits that socially connected users share similarities, our approach is designed to address these challenges. We conduct a thorough evaluation of our method across seven real-world social network datasets, spanning a diverse range of topics and detection tasks, showcasing its applicability to advance research in computational social science.
\end{abstract}

\begin{teaserfigure}
    \centering
    \includegraphics[width=\linewidth]{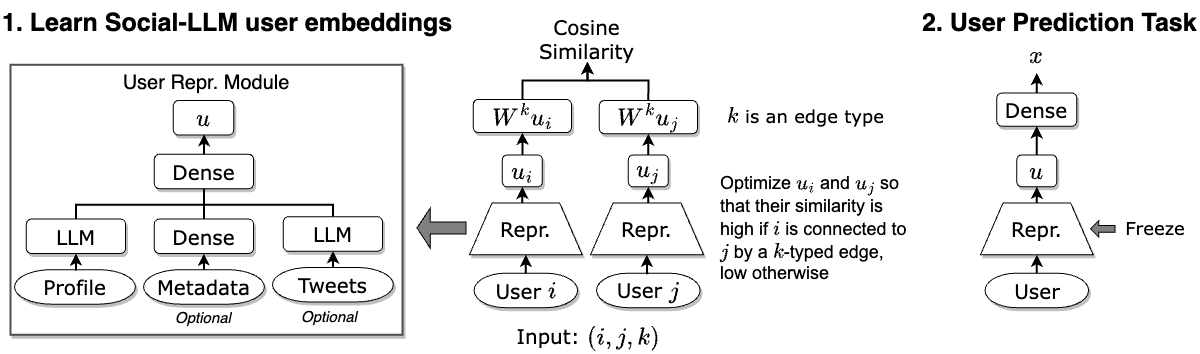}
    \caption{Caption}
    \label{fig:sllm_overview}
\end{teaserfigure}

\maketitle

\section{Introduction}

The surge in popularity of social media over recent decades has provided computational social scientists with an exciting avenue for empirical data-driven mining of human behavior \cite{freeman2004development,lazer2009computational}. Leveraging social network data has enabled the tracking of mass sentiment, health trends, political polarization, the spread of mis/disinformation, social influence, behavior contagion, and information diffusion, all at an unprecedented scale. Social network data comprises two crucial elements: content—\textit{what} people share —and network--who, when, and how frequently users interact with each other. The text-based content aspect of social network data has become more manageable due to recent advancements in large language models (LLMs). However, effectively handling the network element requires methods such as graph representation learning, which often struggle to scale up to large social network data \cite{ma2022graph,wu2021survey}. 

% Yet, the accessibility of big data often surpasses the capabilities of modern technology in terms of hardware resources and processing capabilities \cite{fan2014challenges}. This challenge is particularly pronounced when modeling the network ties of large-scale social networks. While advancements in graph neural networks and graph embeddings have yielded significant progress in network representation, these methods often struggle to mitigate training overhead through conventional approaches like mini-batching or parallelization \cite{serafini2021scalable}. They are also limited in their capacity to generalize to out-of-sample users without retraining. Additionally, social networks, being both larger and sparser compared to other commonly modeled networks (such as protein-protein interaction or citation networks), further intensify the complexity of the problem.

This paper introduces a pragmatic approach for modeling large-scale social network data by leveraging localized social network interactions, building upon the assumption of social network homophily. The social network homophily theory suggests that socially connected users are more likely to be similar \cite{mcpherson2001birds}. Specifically, we harness the power of LLMs by drawing inspiration from the theory of linguistic homophily in social networks, which posits that users with similar language styles are more likely to be friends \cite{kovacs2020language}. Our proposed model, named \textit{Social-LLM}, is influenced by \textit{Retweet-BERT} \cite{retweetbert}, initially designed for political leaning detection using retweet interactions and user profile descriptions. We extend Retweet-BERT by generalizing to all social network interactions, incorporating additional types of content, and applying the model to a diverse range of computational social science applications, including political polarization, online hate speech, account suspension, and morality. % In recent years, there has been no shortage of novel graph embedding or graph neural network methods. However, they all face a critical scalability issue when it comes to large-scale datasets. 
Our summary of contributions is as follows:
\begin{itemize}
\item We propose Social-LLM, a scalable social network representation model that combines user content cues with social network cues for inductive user detection tasks.
\item We conduct a thorough evaluation of Social-LLMs on 7 real-world, large-scale social media datasets of various topics and detection tasks.
\item We showcase the utility of using Social-LLM embeddings for visualization.
\end{itemize}
\section{Related Work}

Detecting users is a crucial element in computational social science research, encompassing a spectrum of areas like identifying ideologies \cite{jiang2020political, barbera2015tweeting}, spotting inauthentic accounts \cite{masood2019spammer, davis2016botornot}, flagging hateful or toxic behavior \cite{jiang2023social, ribeiro2018characterizing}, recognizing influential figures \cite{rios2019semantically}, assessing vulnerability to misinformation \cite{ye2023susceptibility, aral2012identifying}, and beyond.
% ....\todo[inline]{cite} % Existing methods mainly face one of two shortcomings: (1) they don't utilize state-of-the-art results enabled by LLMs, (2) they don't consider the social network data beyond simple network statistical features. We address the first shortcoming by simply including LLMs in our modeling procedure, but that is hardly a novelty
Most user detection methods that utilize social network features only consider them to the extent of network statistics (e.g., node centrality measures) and not the complex relationships among individuals \cite{masood2019spammer,davis2016botornot}. 
% In a survey paper on spam and fake user detection methods, \citet{masood2019spammer} summarized that most methods use simple content features (e.g., number of followers) and network features up to the point of node centrality measures but do not accommodate more complex content features or network ties. Similarly, \citet{davis2016botornot} proposed one of the most highly utilized bot detection models for Twitter, which considers more than 1,000 features but does not harness the full powers of the social network relationship data. Other works similarly compiled user and network features for mental health status detection \cite{saravia2016midas,mcmanus2015mining}.

In many cases, the lack of social network data is due to practical limitations: it's challenging to obtain the network data, and it is also challenging to effectively model them. While the challenge of obtaining network data lies beyond our control, we can address the second limitation with graph representation learning \cite{hamilton2020graph,liu2018heterogeneous}. 
%and GEM is another heterogeneous graph neural network designed for malicious account detection \cite{liu2018heterogeneous}. 
These methods effectively capture crucial higher-order proximity information within social networks yet often demand substantial computational resources during training or have elevated hardware requirements. The scale of large social network data can surpass device capacities that cannot be solved by distributed training, parallelism, or batching due to the inherent inability to partition the graph \cite{serafini2021scalable}. Other methods use sampling approaches to reduce training complexity\cite{ma2022graph}. However, there remains an inherent scalability trade-off when attempting to model very large graphs \cite{wu2021survey}. In this work, we preserve social network data but utilize them in the simplest manner by considering only the first-order proximity (i.e., the edges themselves). We show that such graph approximation is often sufficient for user detection on social media datasets.

Our method, Social-LLM, utilizes multi-relational data among users coupled with user features.  Perhaps the most similar method to ours is TIMME, a scalable end-to-end graph neural network (GNN) user classification method that utilizes multi-relational social network data \cite{xiao2020timme}, or GEM, another heterogeneous GNN designed for malicious account detection \cite{liu2018heterogeneous}. Both methods can also take user content features as node feature input. Because these methods inherently rely on social network relations, it is not possible to deploy them inductively on out-of-sample users. Social-LLM, however, can be applied to any unseen users as long as we have the same user content features without retraining.

% graph is too big to fit on device, even distributed/parallelism doesn't work because you can't partition the graph. recommend sampling 
% \cite{serafini2021scalable}
% \cite{wu2021survey} scalability trade off
\section{Method}

We propose Social-LLM, a model that leverages network homophily and user features to learn user representations scalably. This representation model can then be applied inductively for a range of downstream user detection tasks. Social-LLM draws from two types of social network features: content cues from each user and network cues from social interactions. 

\subsection{Content Cues}
The content cues are derived mainly from the textual content on their social media but can also be from other contextual metadata. We primarily utilize users' profile descriptions, which are self-provided mini-biographies. For most user detection task purposes, users' biography encodes a substantial amount of personal information with personal descriptors (e.g., ``\textit{Mother}'', ``\textit{Senator}'', ``\textit{Research Scientist}) and, in some cases, self-identities and beliefs (e.g., ``\textit{Democratic}'', ``\textit{\#BLM}'', ``\textit{LGBTQ}''). Capped at a limit of 160 characters, these descriptions have to be short, incentivizing users to convey essential information they wish to share about themselves succinctly and attractively. The use of Twitter profile descriptions, not the tweet texts, has proved useful in a large number of computational social science research \cite{jiang2022pronouns,rogers2021using,retweetbert,piao2017inferring,thelwall2021male}. 
% A study also found that the number of followers a user has is correlated with the length of their profiles, suggesting that most influential and popular users have meaningful profile descriptions \cite{mention2018}. 
From a practical standpoint, using user profiles instead of all of the tweets by a user also vastly reduces the complexity of the computation problem as well as alleviates data collection challenges. In addition to profile descriptions, we also leverage, when applicable, user metadata features (e.g., follower counts, account creation date, etc.) and user tweets. 

\subsection{Network Cues}
Online social media platforms offer a variety of ways to engage with one another, such as by following, liking, or re-sharing. These acts of social interaction can be gathered to form social networks. The Twitter API enables us to obtain three types of social interactions: retweeting, mentioning, and following. Though the following network is perhaps the most useful indication of user interaction, it is rarely in empirical research used due to the API rate limits set by Twitter \cite{martha2013study}. As such, following the vast majority of computational social science research on Twitter (e.g., \cite{conover2011political,ferrara2016rise}), we use the retweet and mention networks in this research. \textit{Retweet} refers to the act of re-sharing another user's tweet directly, without any additional comments. \textit{Mention} includes all other acts of mentioning (using `@') another user, including retweeting with comments (i.e., quoting), replying, or otherwise referencing another user in a tweet. We draw a distinction between retweets and mentions because they may represent two distinct motivations of social interaction: retweeting is usually understood as an endorsement \cite{boyd2010tweet,metaxas2015retweets} while mentioning could be used to criticize publicly \cite{hemsley2018tweeting}.

\subsection{Social-LLM Framework}

We train Social-LLM in an unsupervised manner to learn user representations in a $d-$dimensional embedding space (Figure \ref{fig:sllm_overview}, step 1). Once we train the user representation module, we can apply the user representation module to any user content input to obtain the user embeddings. Additional layers can be trained on top of any downstream user detection task (Figure \ref{fig:sllm_overview}, step 2). 
% In step 2, we can freeze the user representation module to reduce training complexity. We omit further explanation of step 2 due to its simplicity. Below, we explain in detail the process of step 1.

\subsubsection{User representation module} 
The user representation module mainly takes in the pre-trained LLM model to be applied to the user's profile description. This LLM model is trainable in order to allow for fine-tuning in our training process. If user metadata features and/or the averaged LLM embeddings of the user tweets are provided, they will be directed through a series of dense layers. We concatenate these outputs into one single embedding and apply another dense layer to produce a single $d$-dimensional embedding $u_i$ for user $i$.

\subsubsection{Unsupervised training via Siamese architecture} 
The user representation module is wrapped in a Siamese model architecture in a manner similar to Sentence-BERT \cite{reimers2019sentence}. 
% which employs identical representation modules on two sentences and optimizes the similarity of their embeddings if the sentences are deemed semantically similar. In our case, 
Specifically, we apply an identical representation module on the user content cue and optimize the resulting embeddings based on the network cues. A training instance of Social-LLM is a tuple $(i, j, k)$ where $i$ and $j$ are two users who are connected by a social network interaction (i.e., an edge) of type $k$. We want to train the user embeddings $u_i$ and $u_j$ so that they are as similar as possible. \citet{reimers2019sentence} and \citet{retweetbert} achieve this by optimizing the cosine similarity of embeddings. However, we also want to consider (1) multiple edge types--modeling retweets distinct from mentions--and (2) directionality--user $A$ retweeting from user $B$ is not the same as user $B$ retweeting from user $A$. To account for multiple edge types, we initialize a learnable weight matrix $W^{k}$ for every edge type $k$. To account for directionality, we can use separate weight matrices $W^{k_{\text{in}}}$ and $W^{k_{\text{out}}}$ for the in- and out-edges. We then calculate the cosine similarity scores between $W^ku_{i}$ and $W^ku_{j}$, or $W^{k_{\text{in}}}u_i$ and $W^{k_{\text{in}}}u_j$ in the directional case, as the final output. We can also account for edge weights by weighting each training instance proportionally to their weight. 

\subsubsection{Multiple negatives ranking loss} 
We train the model with a ranking loss function, pitching positive examples against negative examples. All edges in the graph serve as positive examples, and all other pairs of users, i.e., user $i$ and user $j$ who are \textit{not} connected by an edge, can serve as negative examples. To speed up the training procedure, we use the multiple negatives loss function \cite{henderson2017efficient}, which has shown to work well in \citet{retweetbert}. Essentially, all other pairs of users in the same batch serve as negative examples. For instance, if the input batch contains positive examples $[(i_1, j_1), (i_2, j_2), ...]$, then $\{(i_x, j_y)\}$ for all $x!=y$ are negative examples. This will encourage users who are truly connected in the graph to have more similar representations than users who do not. To minimize training complexity, we alternate the training of different types of edges in a round-robin fashion. For example, if we want to accommodate for both $k=\text{retweet}$ and $k=\text{mention}$ edges, we will train one batch of retweet-only edges, followed by one batch of mention-only edges, and so on.

\subsubsection{Downstream task application}
The Social-LLM model produces reusable user representation that can be used on any downstream user prediction tasks (Figure \ref{fig:sllm_overview}, step 2). We can fine-tune the representation module further or freeze the layers and add task-specific fine-tuning layers on top. We can also append any user-specific features (profile LLM embeddings, user metadata features, etc.) that we used during the Social-LLM training process to the learned Social-LLM user embeddings at this step.

\subsection{Advantages and Disadvantages}
Social-LLM builds on traditional user detection methods by adding social network components. There are two main advantages of Social-LLM over similar GNN approaches.
\begin{itemize}
    \item \textbf{Ease of training}: The time complexity of step 1 is $\mathcal O(|E|)$, and that of step 2 is even quicker at $\mathcal O(|V|)$. Crucially, since we forgo training the complete graph and only focus on edges as if they are independent, we can fit very large datasets via batching.
    \item \textbf{Inductive capabilities}: Since step 2 of the framework no longer relies on the network, we can extend our model to produce embeddings for any users, provided we have their content information, without needing their network information and refitting the whole model. This is called inductive learning, and most graph-embedding approaches either cannot natively support this (e.g., \cite{grover2016node2vec,zhang2019prone}, or they do so at a significantly higher training complexity \cite{hamilton2017inductive}).
    \item \textbf{Reusability}: The Social-LLM embedding training process is separate from the downstream applications so that we can reuse any learned embeddings for different applications.
\end{itemize} 

The advantages of Social-LLM come with costs. Notably, we sacrifice precision and thoroughness for speed and efficiency. Our model focuses only on first-order proximity, or users who are connected immediately by an edge. This undoubtedly loses valuable information from the global network structures or higher-order proximities. However, as we will demonstrate in this paper, in the cases of many user detection problems on social networks, it is \textit{sufficient} to model the localized connections for a \textit{cheap boost} to performance compared to a framework that does not use the social network at all. For these large but sparse real-world social network datasets, the more powerful graph embedding methods may require a lot more training time, memory footprint, or hardware resources for a marginal gain in performance.

\section{Data}

We describe the dataset we use to validate our approach. The first two datasets, \textsc{Covid-Political} and \textsc{Election2020}, were used in Retweet-BERT \cite{retweetbert}, focusing only on using user profile descriptions and retweet interactions to predict political partisanship. 
% For the purpose of being self-contained, we describe the data processing steps done in \citet{retweetbert}. 
To demonstrate the additional capabilities of Social-LLM, we introduce several other datasets that encompass more heterogeneous user metadata and network features. In addition, our new datasets add diversity to the types of labels (partisanship, morality, account suspension, toxicity), prediction methods (classification and regression; single output and multi-output), time spans, and data sizes to demonstrate the robustness of our approach. The summary statistics of all of our datasets are displayed in Table \ref{tab:data_stats}.

\begin{table*}
    \small
    \caption{Summary statistics of our Twitter datasets.}
    \begin{tabular}{lrrrcccrrr}
    \toprule
    & &  \textbf{\# Retweet}&  \textbf{\# Mention} & \textbf{Profile} & \textbf{Metadata} &\textbf{Tweet} & \textbf{Time}&& \textbf{Pred.} \\
    \textbf{Dataset} &\textbf{ \# Users} & \textbf{Edges}  & \textbf{Desc.} & \textbf{Edges} & \textbf{Features} &\textbf{Texts} & \textbf{Span} & \textbf{Label(s)} & \textbf{Type} \\
    \midrule
    \textsc{Covid-Political} & 78,672 & 180,928 & - & \cmark& \cmark& \xmark & 6 Months & Partisanship (1) & Cls.\\
    \textsc{Election2020} &  78,932 & 2,818,603 & - & \cmark& \xmark& \xmark &3 Months &Partisanship (1) & Cls. \\
    \textsc{COVID-Morality} & 119,770 & 609,845 & 639,994 & \cmark& \cmark& \xmark &2 Years &Morality (5) & Reg.\\
    \textsc{Ukr-Rus-Suspended} & 56,440 & 135,053& 255,476& \cmark& \cmark& \cmark &1 Month &Suspension (1) & Cls. \\
    \textsc{Ukr-Rus-Hate} & 82,041 & 166,741 & 414,258 & \cmark& \cmark & \xmark &1 Month&Toxicity (6) &  Reg. \\
    \textsc{Immigration-Hate-08} & 5,759 & 63,097 & 83,870 & \cmark& \cmark & \xmark& All times &Toxicity (5) & Reg.\\
    \textsc{Immigration-Hate-05} &2,188 & 4,827 & 7,993 & \cmark& \cmark & \xmark& All times & Toxicity (5)  &Reg.  \\
    \bottomrule
    \end{tabular}
\label{tab:data_stats}
\end{table*}

\subsection{COVID Politics}\label{sec:data_covid_political}
The COVID-19 pandemic left an unprecedented impact on everyone worldwide. Research has shown that COVID-19 was politicized, with partisanship steering online discourse about the pandemic \cite{jiang2020political,calvillo2020political}. As such, our prediction task is to detect user partisanship from this COVID social network. Our dataset, \textsc{Covid-Politics}, is based on a real-time collection of tweets related to the COVID-19 pandemic \cite{chen2020tracking} between January 21 and July 31, 2020, which was further preprocessed in \citet{retweetbert} to remove bots \cite{davis2016botornot,yang2022botometer}.
% The time period of this dataset is limited to January 21 to July 31, 2020. Following \citet{garimella2018quantifying}, we retain only retweet interactions between users if there are at least two retweet occurrences to filter for stronger indications of retweet endorsement. To eliminate users who were inactive or were not sampled sufficiently to be included in the dataset, we remove users with total degrees (in- and out-degrees) less than 10. Further, to mitigate the influence of social bots, we also removed the top 10\% of users who are most likely to be bots \cite{davis2016botornot,yang2022botometer}. After filtering, 
It also consists of the following user metadata features: initial follower count, final follower count, number of tweets, number of original tweets, number of days active (post at least once), and whether the user is verified. 

The ground truth partisanship labels from \citet{retweetbert} are derived from a blend of two heuristics-labeling methods. The first uses annotated political hashtags used in user profile descriptions, and the second uses the partisanship leaning of new media URLs mentioned in users' tweets.
This dataset contains 78,672 labeled users with 180,928 retweet interactions. The distribution of users is unbalanced, with approximately 75\% of them labeled as left-leaning.

\subsection{Election 2020}
The 2020 US presidential election took place amidst the backdrop of the COVID-19 pandemic. Former Vice President Joe Biden, the Democratic nominee, defeated the incumbent Republican President Donald Trump. The \textsc{Election-2020} dataset is based on a real-time collection of tweets regarding the 2020 US presidential election \cite{chen2021election} from March 1 to May 31, 2020. This dataset was similarly preprocessed to remove bots and users with low degrees \cite{retweetbert}.
% The dataset was also preprocessed in our prior work \cite{retweetbert}, and we retained only the profile description of the users and retweet interactions, not user metadata features. Similar to the \textsc{Covid-Political} dataset (\S\ref{sec:data_covid_political}), we eliminate users who are likely bots (top 10\% of bot scores), edges that occurred only once, users with degrees less than 10. After filtering, we have around 115,000 users and 3.6 million retweet interactions.
Since this is specifically a dataset on US politics, the user label we are interested in predicting is partisanship. We use the same partisanship labels from \citet{retweetbert}, which labeled 78,932 users as either left-leaning or right-leaning. The distribution split is even, with around 50\% of the users labeled as left-leaning. These users encompass the \textsc{Election2020} dataset with 2.8 million retweet interactions.

\subsection{COVID Morality} \label{sec:data_covid_morality}
\begin{sloppypar}

The \textsc{Covid-Morality} dataset is also compiled from the previously mentioned real-time collection of COVID-19 data as \textsc{Covid-Politics} \cite{chen2020tracking}, but spans a longer time period and is focused on a different prediction task. This data spans from February 2020 to  October 2021 for 21 full months. Our task is to predict the moral foundations of users. The Moral Foundation Theory (MFT) decomposes human moral reasoning into five dimensions: care/harm, fairness/cheating, loyalty/betrayal, authority/subversion, purity/degradation  \cite{haidt2004intuitive}.\footnote{A sixth dimension, liberty/oppression \cite{haidt2012righteous}, was later added. However, we only use the first five following \citet{rao2023pandemic}.} 
Research on moral foundation and network homophily found that moral values of purity can predict social network distances between two users, suggesting that social networks exhibit \textit{purity homophily} \cite{dehghani2016purity}. On the subject of COVID-19, numerous studies have established a link between moral values and decision-making regarding health-related behaviors during the pandemic, such as wearing masks or getting vaccinated \cite{chan2021moral,francis2022moral,diaz2022reactance}. It stands to reason that user morality may play a role in facilitating online communication patterns on COVID-19.
\end{sloppypar}
To detect moral values in the tweets, we use the morality detector \cite{rao2023pandemic} fine-tuned specifically for this dataset. It predicts the 10 moralities (each dimension contains two opposite labels for virtues and vices) in a multilabel manner.
% \cite{guo2023data}, trained on fusing multiple annotated moral foundation data sources to detect the 10 main moralities (5 moral dimensions) in a multilabel manner specifically on our dataset \cite{rao2023pandemic}. 
% In particular, this morality detector performs better on our specific dataset compared to other widely used methods . 
We retain tweets that have at least one of the ten moralities present. Of the users who produced these tweets with moral values, we filtered for active users by retaining users who posted at least 10 tweets during any month. We then sampled 150,000 users from this set. With this set of users, we build a network of retweet and mention interactions. As with \citet{retweetbert}, we keep only edges with weights $>=2$, leaving us with 119,770 users sharing 609,845 retweet edges and 639,994 mention edges.  

Our prediction task is a multi-output regression problem. Due to the way the moral datasets are typically annotated, a tweet could be labeled as having any of the 10 morality, even if we don't need to differentiate virtues and vices for the purposes of analyzing moral foundations \cite{hoover2020moral}. For example, a tweet labeled as having care but no harm, harm or no care, or both care and harm all reflect a moral foundation in the care/harm dimension. Therefore, for each tweet, we aggregate the 10 labels into 5, assigning a value of 1 to the moral foundation if both polar opposites are present, 0.5 if either the virtue or vice is present, and 0 otherwise. We then calculate 5 moral scores for users by computing the average moral foundation score across all of their tweets. 
% The distributions of the labels are shown in Figure \ref{fig:covid_mf_label_dist}.

This dataset includes the following user metadata features: account age,  number of followers count, number of people they are following, number of lists the user is a member of, the total number of tweets ever posted by this user, the number of posts favorited by the user, and whether they are verified. Additionally, we calculate the number of original tweets, retweets, quoted tweets, and replies by each user contained in our dataset.

\subsection{Ukraine-Russia Suspended Accounts}\label{sec:data_ukr_rus_suspended}

When the Ukraine-Russian war erupted in early 2022, social media quickly became a platform to spread content about the conflict, not all of which are truthful. Early research suggests that Russian mis/disinformation campaigns, state-sponsored content, and otherwise suspicious activities were rampant on social media \cite{pierri2023propaganda,indiana2022suspicious}. We use a real-time collection of tweets about the conflict \cite{chen2023ukrrus} for a full month in March 2022. Many of these users who tweeted about the war were since suspended by Twitter,\footnote{See Twitter's help page on account suspension: \url{https://help.twitter.com/en/managing-your-account/suspended-x-accounts}} and they were often found to be newer, more active (spamming), and more toxic than users that are not suspended \cite{pierri2023does}. We theorize that suspended users and normal users reflect different communication characteristics due to different agendas, motivations, and needs from social media.

Our task is to predict whether a user was eventually suspended from their user metadata and network features. Our raw dataset contains around 10 million non-suspended users and 1 million suspended users. We filter for users who posted a minimum of 10 tweets to remove inactive users and a maximum of 130 tweets to remove spamming bots. 
% Although bots could certainly also be used to advance disingenuous agendas, we want to focus on accounts that are more likely to be controlled by real humans. 
At this time of the research, we no longer had access to the Botometer \cite{davis2016botornot} tool due to Twitter's API cutoff \cite{botometershutdown}. Therefore, we resort to using the maximum tweet amount as a rough elimination criterion. 130 is the 90\% threshold of the number of tweets per user after removing users with less than 10 tweets. Having 130 tweets in one month represents 4.19 tweets per day, which we feel is a reasonable number of tweets an authentic account could post. After filtering, we have 1.4 million non-suspended and 73,000 suspended users. Since the dataset is heavily imbalanced, we sampled a roughly equal proportion of users who were not suspended. We then built the retweet and mention network using the edge weight $>=2$ criteria and removed any users who were isolated in the network following \cite{retweetbert}. Our final \textsc{Ukr-Rus-Suspended} dataset consists of 56,440 users, 135,053 retweet edges, and 255,476 mention edges. Around 58\% of the users were suspended. We also retained the same user metadata features as we did in the \textsc{Covid-Morality} dataset (\S\ref{sec:data_covid_morality}). Unlike other datasets, the ground-truth labels were not derived from the tweet texts themselves, so we additionally explore including the tweet texts as a content feature in our model.

\subsection{Ukraine-Russia Hate}
Beyond misinformation and coordinated activity, the online discourse regarding the Ukraine-Russia war is also riddled with toxic language \cite{thapa2022multi}. As a spin-off from the previous \textsc{Ukr-Rus-Suspended} dataset (\S\ref{sec:data_ukr_rus_suspended}), we experiment with whether we can detect users' toxicity levels from their Twitter behavior and activity. Our preprocessing step diverges from the \textsc{Ukr-Rus-Suspended} dataset (\S\ref{sec:data_ukr_rus_suspended}) after filtering for users based on the min and max number of tweets.

We employ the Perspective API,\footnote{\url{https://perspectiveapi.com/}} a widely used toxicity detector used in similar studies \cite{frimer2023incivility,kim2021distorting}. The Perspective API returns a  \texttt{TOXICITY} score for each text on a scale of 0 (not toxic) to 1 (very toxic). Besides the flagship \texttt{TOXICITY} score, the API also computes 5 other toxicity measures: \texttt{IDENTITY\_ATTACK}, \texttt{INSULT}, \textsc{PROFANITY}, \texttt{THREAT}, and \texttt{SEVERE\_TOXICITY}. We apply the Perspective API to users' original tweets. Occasionally when the tweet is not written in an unsupported language or contains only URLs, the Perspective API will fail to produce toxicity scores. Therefore, we filter for users who have at least 10 original tweets rated by the Perspective API. There are 82,041 users in this \textsc{Ukr-Rus-Hate} dataset. We save all 166,741 retweet edges and 414,258 edges, neglecting to remove the edges with weights $<2$ as we did previously since the density of the network is comparatively smaller. 
% The distribution of and correlation among the toxicity labels are presented in Figure \ref{fig:ukr_rus_hate_label_dist} and Figure \ref{fig:ukr_rus_hate_label_corr}, respectively. 
The user metadata features we retain are the same as we did in the \textsc{Ukr-Rus-Suspended} (\S\ref{sec:data_ukr_rus_suspended}) and \textsc{Covid-Morality} (\S\ref{sec:data_covid_morality}) datasets.

\begin{figure}
    \centering
    \includegraphics[width=\linewidth]{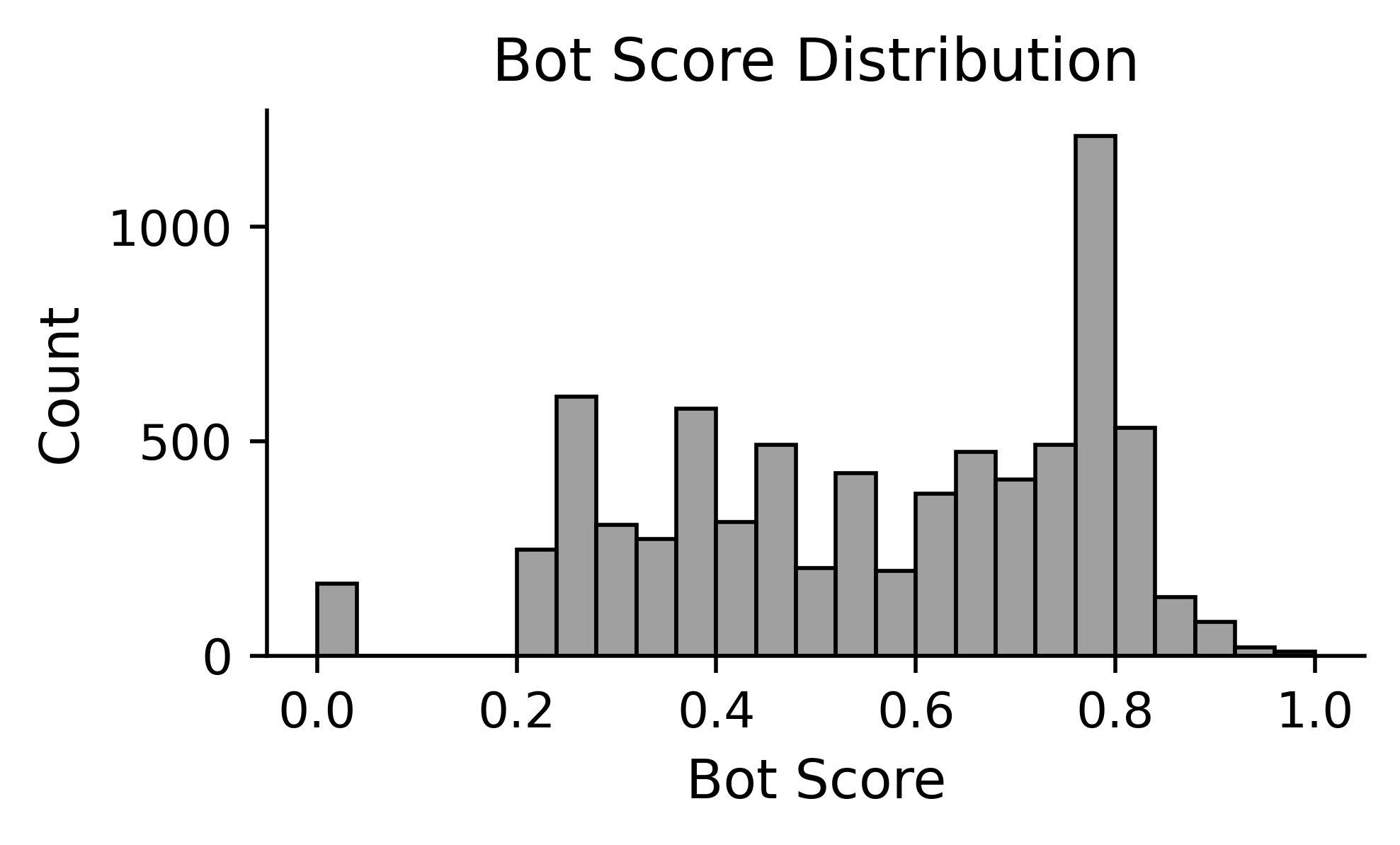}
    \caption{Distribution of user bot scores prior to user preprocessing for the \textsc{Immigration-Hate} datasets.}
    \label{fig:hate_bot_dist}
\end{figure}

\subsection{Immigration Hate}
We compile another hate speech dataset, this time forgoing breadth for depth by collecting a relatively full set of historical tweets by a smaller set of users. This dataset is based on another dataset collected by \citet{bianchi2022njh}, which consists of annotated tweets from 2020-2021 that reference immigration hate terms. We attempted to rehydrate the 18,803 tweets that were found to be uncivil, intolerant, or both. 8,790 (47\% of the total) tweets by 7,566 unique users were successfully retrieved; the rest of the tweets were no longer available. Since these users are known to have tweeted hateful immigration tweets at least at some point, we use them to snowball our \textsc{Immigration-Hate} dataset with the Twitter historical API, collecting the hateful users' most recent tweets up to a maximum of 3,200. This resulted in 21 million tweets, of which 2.9 million tweets were original. We focus only on original tweets in this work.

We apply the Perspective API on the tweets for five measures of toxicity:  \texttt{TOXICITY}, \texttt{IDENTITY\_ATTACK}, \texttt{INSULT}, \textsc{PROFANITY}, and \texttt{THREAT}, then we aggregate toxicity scores for users by computing the average toxicity score of their tweets. Using Botometer \cite{davis2016botornot,yang2022botometer}, we found that bots seem especially prevalent in the dataset (Fig. \ref{fig:hate_bot_dist}). Therefore, to mitigate the influence of bots, we remove users according to two thresholds of bot score: 0.8, which is a conservative choice given the peak in the distribution of bot scores, and 0.5, which would leave us substantially fewer users but with a higher certainty that they are genuine. Since the number of users is already small, we retain all network edges. In the \textsc{Immigration-Hate-08} dataset, 5,759 users with bot scores less than or equal to 0.8 share 63,097 retweet and 83,870 mention edges. In the \textsc{Immigration-Hate-05} dataset, 2,188 users with bot scores less than or equal to 0.5 share 4,827 retweet and 7,993 mention edges. 
% We display the distribution of the users' toxicity scores in Figure \ref{fig:hate_label_dist} and the correlations among the toxicity labels in Figure \ref{fig:hate_label_corr}. 
The user metadata features we use include account age, whether they are verified, number of followers, number of followings, number of total tweets posted, and the number of lists they are a member of.

\section{Evaluation}
We evaluate our method in step 2 of our overall workflow (Figure \ref{fig:sllm_overview}). We conduct an extensive comparison of Social-LLM with baseline methods, as well as several sensitivity and ablation studies.

\subsection{Baseline Methods}
For a thorough evaluation of our approach, we use a series of state-of-the-art baseline methods divided into three camps: content-based, network-based, and hybrid methods. The content-based and network-based models provide an alternative user embedding that we can utilize in the evaluation procedure (Figure \ref{fig:sllm_overview} step 2). All input embeddings undergo similar training processes for target task prediction. For the hybrid method, we use TIMME, an end-to-end user detection method that also uses both user features and network features. We conduct a thorough hyperparameter tuning process for all of the baseline models.

\subsubsection{Content-Based Methods}
\begin{sloppypar}

For Content-Based Methods, we primarily investigate using embeddings from pre-trained LLMs. Fine-tuning the LLMs for our specific purpose is also one option; however, doing so on the \textsc{Covid-Politics} and \textsc{Election2020} dataset did not deliver a substantial enough improvement to justify the added training cost. In this work, we experiment with the following three LLMs applied to the profile descriptions: (1) RoBERTa \cite{liu2019roberta} (\texttt{roberta-base}), (2) BERTweet \cite{bertweet} (\texttt{vinai/bertweet-base}), a RoBERTa fine-tuned on Twitter data, and (3) SBERT-MPNet (\texttt{sentence-transformers/all-mpnet-base-v2}), a Sentence-BERT \cite{reimers2019sentence} model based on MPNet \cite{song2020mpnet} and is currently the best-performing Sentence-BERT model.\footnote{\url{https://www.sbert.net/docs/pretrained_models.html} (Accessed 2023-11-10).}
\end{sloppypar}
For datasets with additional metadata features, we also experiment with using only the raw metadata features as the ``user embeddings'' as well as with concatenating the LLM embeddings with the raw metadata features. For \textsc{Ukr-Rus-Suspended}, we additionally experiment with applying the aforementioned three LLMs on users' tweets, averaging one LLM embedding per user.

\subsubsection{Network-Based Methods}
We use two purely network-based methods as the network-based baseline: node2vec \cite{grover2016node2vec} and ProNE \cite{zhang2019prone}. While GraphSAGE \cite{hamilton2017inductive} is another obvious choice for inductive graph representation learning with node attributes, it is difficult to train on a large graph within reasonable time limits and can therefore underperform \cite{retweetbert}. These network embedding methods do support weights and directions but heterogenous edge types. Therefore, we run a separate network model on the (1) retweet edges only, (2) mention edges only, and (3) indiscriminately combining retweet and mention edges as one edge (with edge weights equal to the sum of the retweet edge and mention edge weights).

\subsubsection{Hybrid Method}
We use TIMME as our hybrid method baseline, providing it with the user content features and network features as our Social-LLM model. The original model was only designed for user classification tasks, but we modified the open-sourced code to enable regression. As TIMME is primarily advertised as a multi-relational model, we mainly apply it on both retweet and mention edges, but we also experiment with combining these edges indiscriminately. 

\subsection{Experimental Setup}
For every dataset and its corresponding set of user embeddings, we conduct the same train-test procedure repeated 10 times, splitting the dataset randomly using 10 pre-selected random seeds. Unless otherwise indicated, we use a 60\%-20\%-20\% train-val-test split. The validation sets are used for early stopping and for model selection. The model architecture and hyperparameters are fixed for all experiments. The classification tasks are evaluated using Macro-F1, and the regression tasks are evaluated using Pearson's correlation. For regression tasks with multiple labels, we average the Pearson's correlation across labels.

\begin{table*}[]
\centering
\small
\caption{Results of Social-LLM on various datasets compared to baseline models. The best model is in bold, and the best baseline model is underlined.}
\label{tab:results}
\begin{tabular}{@{}llccwr{1.5cm}wr{1.5cm}wr{1.5cm}wr{1.5cm}wr{1.5cm}wr{1.5cm}wr{1.5cm}}
% \begin{tabular}{@{}llccrrrrrrr@{}}
\toprule
% & \multicolumn{3}{c}{\textbf{Classification} (Metric: Macro-F1)} & \multicolumn{4}{c}{\textbf{Regression} (Metric: Pearson)} \\ 
% \cmidrule(lr){2-4} \cmidrule(lr){5-8}
% & & & &\textsc{Election} & \textsc{Covid-} & \textsc{Ukr-Rus-} & \textsc{Covid-} & \textsc{Ukr-Rus-} & \textsc{Immigration-} & \textsc{Immigration-} \\
% & & & & \textsc{2020} & \textsc{Political} & \textsc{Suspended} & \textsc{Morality} & \textsc{Hate} & \textsc{Hate-05} & \textsc{Hate-08} \\
& & & &\textsc{Election} & \textsc{Covid-} & \textsc{Ukr-Rus-} & \textsc{Covid-} & \textsc{Ukr-Rus-} & \multicolumn{2}{r}{\textsc{Immigration-Hate-}} \\
& & & & \textsc{2020} & \textsc{Political} & \textsc{Suspended} & \textsc{Morality} & \textsc{Hate} & \textsc{05} & \textsc{08} \\
\cmidrule(lr){5-7} \cmidrule(lr){8-11}
& & \textbf{C} & \textbf{N} & \multicolumn{3}{c}{Cls. (Metric: Macro-F1)} & \multicolumn{4}{c}{Reg. (Metric: Pearson)} \\ 
\midrule
\multicolumn{10}{l}{\textit{Experiment 1: LLMs}}\\
& RoBERTa & \cmark & \xmark & 80.11 & 78.41 & 56.21 & 32.84 & 36.54 & 12.06 & 9.30 \\
& BERTweet & \cmark & \xmark  & 79.31 & 78.42 & 55.69 & 30.72 & 40.38 & 14.33 & 12.03 \\
& SBERT-MPNet & \cmark & \xmark  & \textbf{86.47} & \textbf{82.99} & \textbf{56.79} & \textbf{36.77} & \textbf{43.35} & \textbf{17.16} & \textbf{16.76 }\\
\midrule
\multicolumn{10}{l}{\textit{Experiment 2 (Main): Baselines vs Social-LLM}}\\
& (a) Profile LLM  & \cmark & \xmark & \underline{86.47} & 82.99 & 56.79 & 36.77 & 43.35 & 17.16 & 16.76 \\
& (a) + (b) Metadata  & \cmark & \xmark & - & 83.26 & 70.75 & 40.43 &\underline{ 45.38} & 17.72 & 17.32 \\
& (a) + (b) + (c) Tweet LLMs & \cmark & \xmark & - & - & \underline{81.74} & - &- & - &-  \\
& (d) node2vec  & \xmark & \cmark & - & \underline{88.65} & 72.33 & 50.53 & 39.97 & 10.70 & 12.18 \\
& (e) ProNE  & \xmark & \cmark &76.28 & 64.04 & 77.95 & \underline{\textbf{51.13}} & \underline{45.38} & 5.47 & 14.30 \\
& (f) TIMME & \cmark & \cmark & 84.81 & 81.85 & 72.91 & 30.47 & 43.46 & \underline{20.98} & \underline{18.67}\\
% & \textbf{Social-LLM} & \cmark & \cmark& \textbf{97.87} & \textbf{90.82} & \textbf{78.41} & 50.15 & \textbf{57.27} & \textbf{21.17} & \textbf{20.11} \\

% & Profile LLM  & \cmark & \xmark & 86.47 & 82.99 & 56.79 & 36.77 & 43.35 & 17.16 & 16.76 \\
% & Profile LLM + Metadata  & \cmark & \xmark & - & 83.26 & 70.75 & 40.43 & 45.38 & 17.72 & 17.32 \\
% & node2vec  & \xmark & \cmark & - & {88.65} & 72.33 & 50.53 & 39.97 & 10.70 & 12.18 \\
% & ProNE  & \xmark & \cmark &76.28 & 64.04 & {77.95} & \textbf{51.13} & {45.38} & 5.47 & 14.30 \\
& \textbf{Social-LLM} & \cmark & \cmark& \textbf{97.87} & \textbf{90.82} & \textbf{82.71} & 50.15 & \textbf{57.27} & \textbf{21.17} & \textbf{20.11} \\
% \midrule
& \%$\uparrow$ && & 13\% & 2\% & 1\% & -2\% & 26\% & 1\% & 7\% \\
\midrule
% \multicolumn{10}{l}{\textit{Experiment 2: Ensemble Models}}\\
% & (a) + (b) + (c) + \{(d) or (e)\} & \cmark & \cmark & 90.40 & 90.30 & 80.31 & 54.14 & 54.84 & 20.41 & 18.69\\
% % & \textbf{Social-LLM} + (c) & \cmark & \cmark& - & \textbf{91.75} & \textbf{82.67} & \textbf{55.85} & \textbf{58.79} & 20.87 & \textbf{20.81} \\
% % & \textbf{Social-LLM} + (d) & \cmark & \cmark & 97.85 & 90.86 & 82.51 & 56.65 & 58.53 & 21.52 & 20.34 \\
% & \textbf{Social-LLM} + \{(d) or (e)\} & \cmark & \cmark& \textbf{97.85} & \textbf{91.75} & \textbf{82.67} & \textbf{55.85} & \textbf{58.79} & \textbf{20.87} & \textbf{20.81} \\
% & \%$\uparrow$ & && 8\% & 2\% & 1\% & 5\% & 11\% & 5\% & 11\% \\
% \midrule
\multicolumn{10}{l}{\textit{Experiment 3: Ablation on edge types in Social-LLM models}}\\
& RT & \cmark & \cmark & - & - & 70.71 & 46.57 & 48.18 & 18.85  & 18.18 \\
& MN & \cmark & \cmark & - & - &71.32 & 45.33  & 49.55 & 18.73  & 17.92 \\
& RT \& MN (distinct) & \cmark & \cmark & - & - &71.99 & 20.40  & 51.20  & 14.75 & 18.89  \\
& RT + MN (indistinct) & \cmark & \cmark & - & - & \textbf{72.10} & \textbf{47.51} & \textbf{50.73}  & \textbf{19.05} & \textbf{18.53}  \\
\midrule
\multicolumn{10}{l}{\textit{Experiment 4: Ablation on edge directions and weights in Social-LLM models}}\\
& (best edge combo model) & \cmark & \cmark & 97.78 & 90.68  & 72.10 & 47.51 & 50.73 & 18.53  & 19.05  \\
&  + w & \cmark & \cmark & 97.78& 90.55  & 71.85 & 46.98 & 51.46 & 18.70  & 18.81  \\
&  + d & \cmark & \cmark & \textbf{97.85} & \textbf{90.82}& 71.77& \textbf{50.15}  & \textbf{57.19} & \textbf{18.95}  & 18.77  \\
&  + d + w & \cmark & \cmark & 97.82 & 90.42 & \textbf{72.17} & 46.89 & 49.35 & 17.67 & \textbf{19.21}\\
\bottomrule
\end{tabular}
\end{table*}

\section{Results}
Below, we discuss our experimental results and interpret our findings. For model selection within the same family of methods, we use the validation sets to select the final model. Most results are presented in Table \ref{tab:results}, which shows the average result over 10 repeated random train-test splits. 

\subsection{Experiment 1: Choice of LLMs}
We first experiment with the choice of LLMs, which determines both the best baseline method for LLMs and which LLM we should use in our Social-LLM models. We selected the clear winner, SBERT-MPNet, which outperformed RoBERTa and BERTweet on all datasets. We note that, given rapid innovations in NLP and LLMs, SBERT-MPNet may not be the best model or could soon be replaced by a better successor. However, the contribution of Social-LLM is not tied to a single LLM but rather a model training paradigm that can be paired with any LLM. Our choices of LLMs are driven by ease of use, costs, and reproducibility \cite{bosley2023we}. 

\begin{figure}
    \centering
    \includegraphics[width=0.9\linewidth]{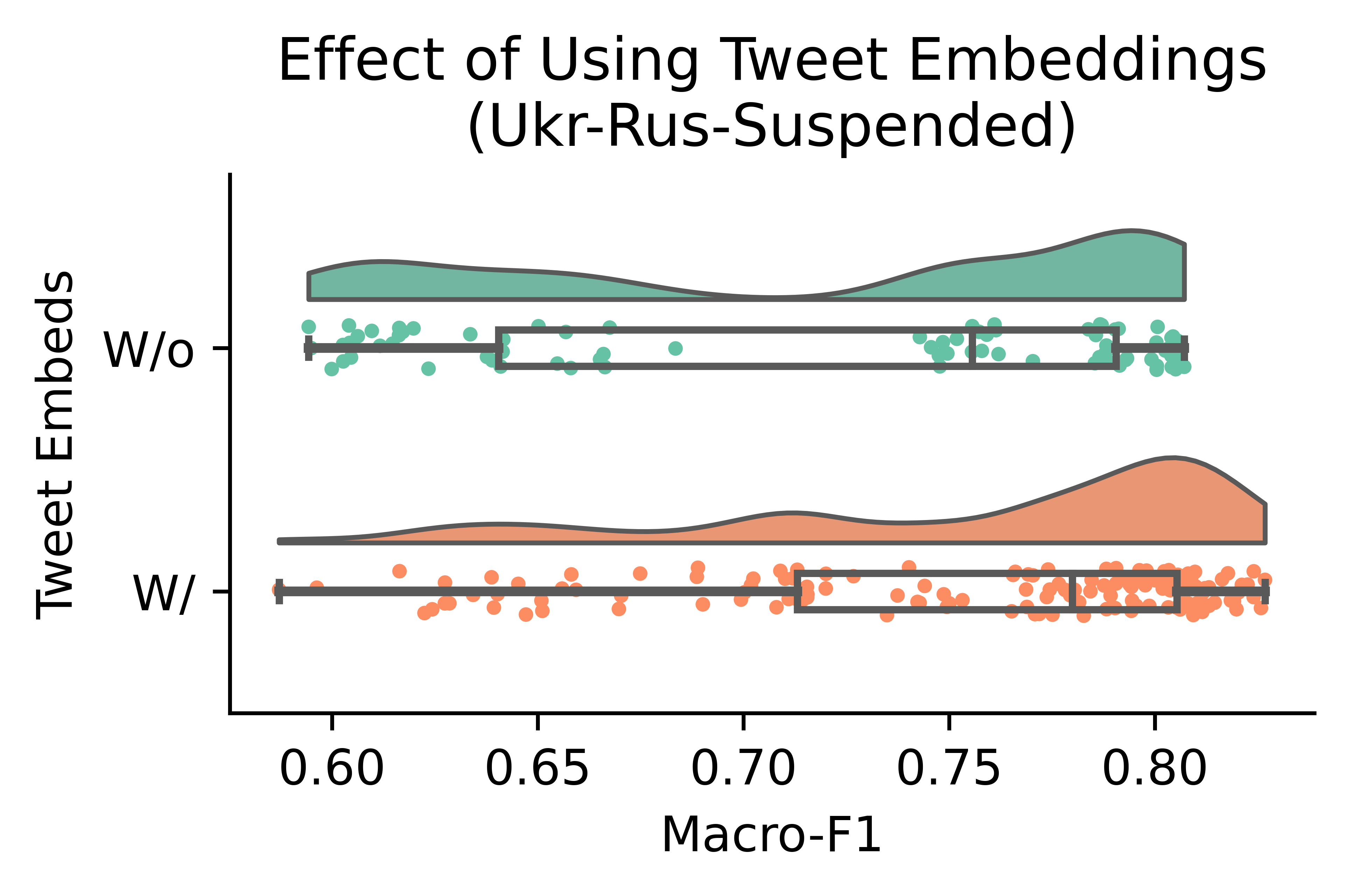}
    \caption{Ablation study of user tweet embeddings on the \textsc{Ukr-Rus-Suspended} dataset (Experiment 5).}
    \label{fig:use_tweet_embeds_ukr_rus_suspended}
\end{figure}
\begin{figure*}
    \centering
    \includegraphics[width=\linewidth]{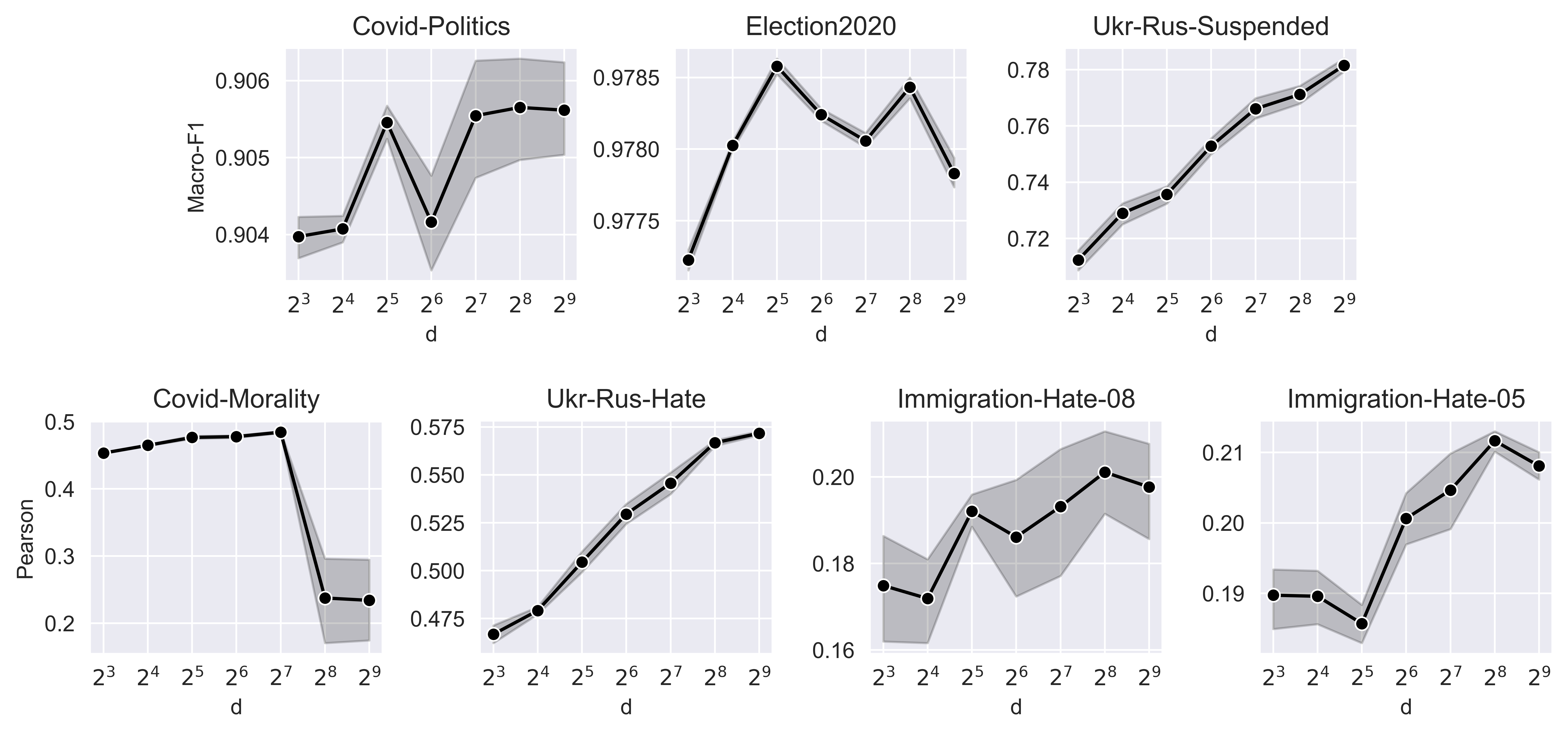}
    \caption{Sensitivity to embedding dimension $d$ (Experiment 6).}
    \label{fig:exp_dimension}
\end{figure*}

% \begin{figure*}
%     \centering
%     \includegraphics[width=\linewidth]{figures/exp_training_size.png}
%     \caption{Sensitivity to training sizes (Experiment 7).}
%     \label{fig:exp_training_size}
% \end{figure*}
\subsection{Experiment 2: Main Experiments}
In Table \ref{tab:results} Experiment 2, we underline the best baseline model and boldface the best model (either a baseline or the Social-LLM model) for each dataset. We also indicate the percentage gain or loss in performance using Social-LLM compared to the best-performing baseline. Regarding baseline methods, there is no clear winner among the content-based, network-based, or hybrid models. Network-based models face much higher variability in performance across datasets, pointing at issues when using solely network features. Notably, we observe that Social-LLM is superior in nearly all cases, with improvements ranging from a substantial 26\% to a modest 1\%. Using a one-sided $t$-test, we find that all improvements are statistically significant. The only dataset where Social-LLM performs comparatively worse than the baselines is \textsc{Covid-Morality}, where the network embedding models are slightly superior, but the Social-LLM model still demonstrates commendable performance. In summary, Social-LLM emerges as the most consistent model, exhibiting robustness across various tasks and data sizes.

% \subsection{Experiment 2: Ensemble Models}

\subsection{Experiment 3: Edge Type Ablation}
We perform an edge-type ablation experiment to evaluate the importance of each edge type on datasets containing both retweets and mention edges. When using only one type of edge, we find they perform comparably. The combination of retweets and mentions as two distinct edge types occasionally results in improved performance but can also lead to deteriorated outcomes. However, combining them indiscriminately as one edge type yields the best performance consistently. This suggests that both retweets and mentions carry important signals, yet the distinctions between the two actions might not be substantial enough to warrant differentiation for our objective tasks.

\subsection{Experiment 4: Edge Weights and Directions}
Using the best edge-type model (RT for \textsc{Election2020} and \textsc{Covid-Political}, and RT + MN for all others), we then experiment with adding edge weights (+ w) and edge directions (+ d). The inclusion of directions always yields better performance, and occasionally, the performance is further enhanced if we stack on weights as well. The importance of directionality emphasizes the value of understanding the flow of information exchange on social networks.

\begin{figure*}
    \centering
    \includegraphics[width=0.3\linewidth]{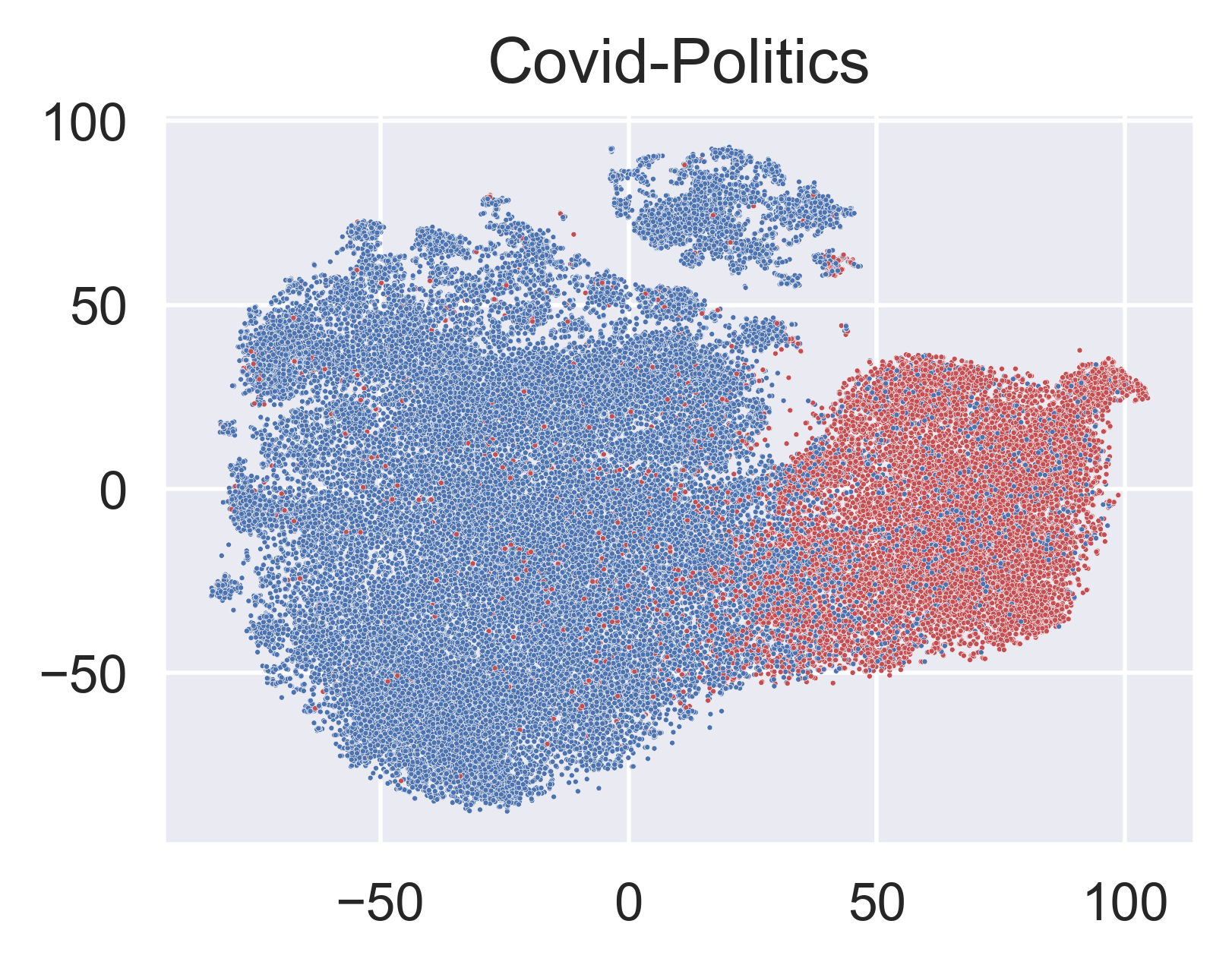}
    \includegraphics[width=0.3\linewidth]{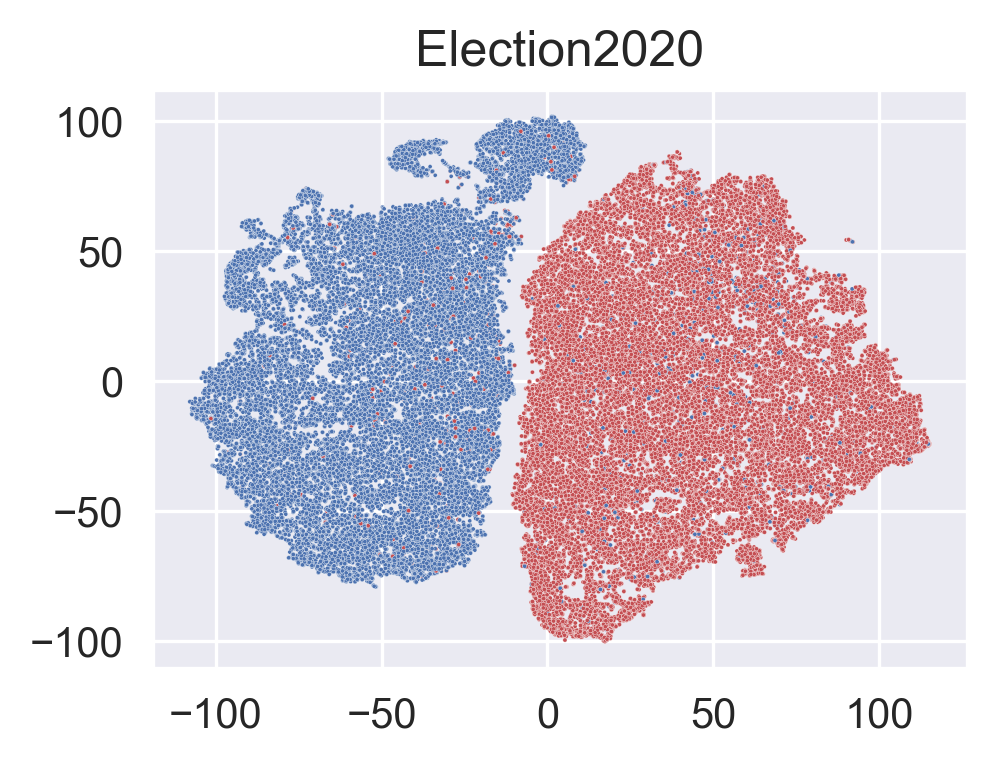}
    \includegraphics[width=0.3\linewidth]{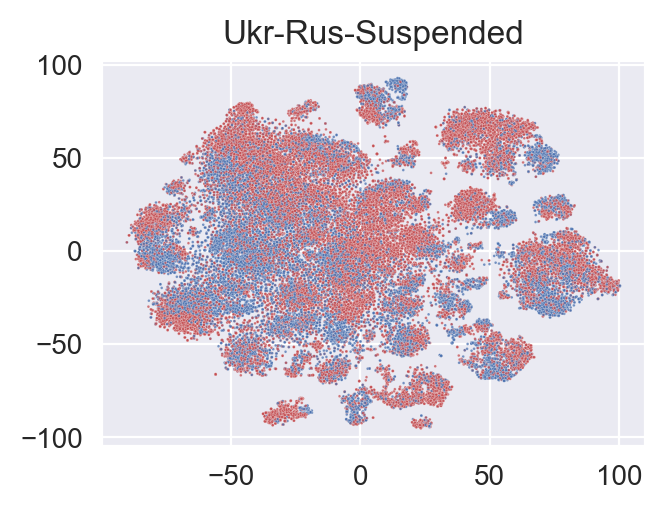}
    \caption{Visualization of Social-LLM embeddings on select datasets.}
    \label{fig:viz}
\end{figure*}

\subsection{Experiment 5: User Tweet Embeddings}
\textsc{Ukr-Rus-Suspended} is the only dataset for which the ground truth labels were not derived directly from the tweet texts; therefore, we can additionally include user tweet embeddings as features. In Fig. \ref{fig:use_tweet_embeds_ukr_rus_suspended}, we see that using user tweet embeddings leads to an average improvement of 4\% Macro-F1 between otherwise identically configured models. This experiment underscores the importance of including user tweets, when applicable and suitable, in user prediction tasks.

\subsection{Experiment 6: Sensitivity to Dimension Size}
For every dataset, we select the Social-LLM model with the best edge type, edge weight, and edge direction configuration to plot the sensitivity to embedding dimension $d$. The results are presented in Figure \ref{fig:exp_dimension}. Performance generally increases with rising dimensions, with $d=258$ being a popular choice; however, we note that Social-LLM usually performs quite well even with very low dimensions.

% \subsection{Experiment 7: Sensitivity to Training Size}
% We compare how Social-LLM fares with baseline methods under varying conditions of training size. On \textsc{Covid-Political}, \textsc{Election2020}, and \textsc{Ukr-Rus-Hate}, Social-LLM consistently yields higher performance even given a very small training size. In other datasets, Social-LLM demonstrates similar or slightly inferior performance compared to the best-performing baseline but is never substantially worse than the best baseline. Given the variability in the baseline models across datasets, the consistency and robustness in Social-LLM performance is noteworthy. 

\subsection{Visualization}
Finally, we highlight the utility of using Social-LLM embeddings as visualization tools. In Figure \ref{fig:viz}, we use TSNE \cite{tsne} to reduce the embedding dimension from selected datasets and visually represent them. We observe a distinct separation of liberals and conservatives in \textsc{Covid-Politics}, and this distinction is even pronounced in \textsc{Election2020}, emphasizing the political differences among users on a more politics-oriented topic. In the \textsc{Ukr-Rus-Suspended} dataset, the global separation between suspended and non-suspended accounts is less apparent, but localized clusters of suspended and non-suspended users emerge. In sum, Social-LLM embeddings offer valuable support in visualizing complex social networks. 

\section{Conclusion}
This paper presents Social-LLM, a scalable social network embedding method that integrates user content information, primarily from user profile descriptions, with social interaction information (e.g., retweets and mentions) for user detection. By combining the state-of-the-art innovations in LLMs with straightforward modeling of first-order proximity---specifically, considering only the edges themselves---from the social network by exploiting characteristics of social network sparsity and homophily. Leveraging 7 different large Twitter datasets drawn from the real world, with a diverse range of meaningful user detection tasks, we showcase the advantage and robustness of our method against state-of-the-art methods that rely solely on content or network features. Importantly, Social-LLM, once fitted to the social network, can be applied to numerous downstream user prediction tasks even \textit{in the absence} of the original social network, underscoring its efficiency and generalizability to out-of-sample users. We show that Social-LLM works best when modeling both retweet and mention edges indiscriminately and when accounting for the directionality of social network interactions. Using additional tweet content embeddings also improves the performance. Further, Social-LLM embeddings prove useful when visualization large-scale social networks.
% \section*{Appendix}

\bibliographystyle{ACM-Reference-Format}
\bibliography{main}

\end{document}